\documentclass[prb,twocolumn,amssymb,amsmath,superscriptaddress,floatfix]{revtex4-1}
\usepackage{graphicx}
\usepackage{pstricks}
\usepackage{amsmath}

\begin{document}

\title{Unified description of ground and excited states  
of finite systems: the self-consistent \textit{GW} approach}
\author{F.~Caruso}
\email{caruso@fhi-berlin.mpg.de}
\author{P.~Rinke}
\author{X.~Ren}
\affiliation{Fritz-Haber-Institut der Max-Planck-Gesellschaft, Faradayweg 4-6, D-14195 Berlin, Germany} 
\affiliation{European Theoretical Spectroscopy Facility}
\author{M.~Scheffler} 
\affiliation{Fritz-Haber-Institut der Max-Planck-Gesellschaft, Faradayweg 4-6, D-14195 Berlin, Germany}
\affiliation{European Theoretical Spectroscopy Facility}
\author{A.~Rubio}
\affiliation{Fritz-Haber-Institut der Max-Planck-Gesellschaft, Faradayweg 4-6, D-14195 Berlin, Germany} 
\affiliation{European Theoretical Spectroscopy Facility}
\affiliation{Nano-Bio Spectroscopy group and ETSF Scientific Development Centre, 
Universidad del Pa\'is Vasco, CFM CSIC-UPV/EHU-MPC and DIPC, Av.\ Tolosa 72, E-20018 Donostia, Spain}
\begin{abstract}
$GW$ calculations with fully self-consistent Green function $G$ and
screened interaction  $W$ -- based on the iterative 
solution of the Dyson equation -- 
provide a consistent framework for the description of ground and excited state 
properties of interacting many-body systems.
We show that for closed shell systems self-consistent $GW$ reaches the same final
Green function regardless of the initial reference state. Self-consistency systematically improves ionization
energies and total energies of closed shell systems compared to
$G_{\rm 0}W_{\rm 0}$ based on Hartree-Fock and (semi)local density-functional theory. 
These improvements also translate to the electron density as exemplified by an improved description of dipole moments and permit us to assess the quality of ground state properties such as  bond lengths and vibrational frequencies.
\end{abstract}
\maketitle

Many-body perturbation theory (MBPT) \cite{fetter}
in the $GW$  approximation of the electronic self-energy \cite{Hedin1965,thegwmethod}
is presently the state-of-the-art method for the 
description of the spectral properties of solids.
\cite{Aulbur/Jonsson/Wilkins:2000,Onida/Reining/Rubio} Recently, it has
steadily gained popularity for molecules and nanosystems.\cite{thygesen/transport}
In addition MBPT provides a prescription to extract total 
energies and structural properties from the $GW$ approximation and therefore
a {\it consistent theoretical framework
for single-particle spectra and total energies}.

Due to its numerical cost and algorithmic difficulties, 
the $GW$ method has only recently been applied
self-consistently (i.e. non-perturbatively) 
to atoms,\cite{stan} molecules \cite{thygesen} and 
molecular transport.\cite{thygesen/transport}
Predominantly, $GW$ calculations are still performed 
perturbatively ({\it one-shot} $G_{\rm 0}W_{\rm 0}$)
on a set of single particle orbitals and eigenvalues obtained
from a preceding density functional theory \cite{kohnsham1965}
(DFT) or Hartree-Fock (HF) calculation.
This procedure introduces a considerable 
starting point dependence,
\cite{patrick2005/all,Fuchs/etal:2007/all,patrickpssb/all} which  
can be eliminated by iterating the Dyson equation to self-consistency.\cite{holmvonbarth1998,stan,thygesen,thygesen/transport}
The resulting self-consistent $GW$ (sc-$GW$) framework is a 
conserving approximation in the sense of Baym and Kadanoff 
\cite{baymkadanoff} (i.e. it satisfies momentum, energy, 
and particle number conservation laws). 
sc-$GW$ gives total energies \cite{Almbladh/etal:1999} free 
from the ambiguities of the $G_{\rm 0}W_{\rm 0}$ scheme, in which
the results depend on the chosen total energy functional. \cite{stan}
However, as in any 
self-consistent theory, the question remains 
if the self-consistent solution of the Dyson equation is unique.
This issue is fundamentally different from the 
initial-state dependence of $G_{\rm 0}W_{\rm 0}$.  For HF 
\cite{Thom/Head-Gordon:2008} and LDA/GGA+$U$ \cite{Meredig10/all}  
calculations, it is well known that the self-consistency 
cycle can reach many local minima instead of the global minimum. 
Moreover, a previous sc-$GW$ study for the Be atom showed 
that norm-conserving pseudopotential calculations do not produce the same 
final $GW$ Green function (and corresponding ionization 
potential) as all-electron calculations.\cite{Delaney/etal:2004/all}

In this communication, we demonstrate certain key aspects 
of the sc-$GW$ approximation for closed shell molecules, 
that make sc-$GW$ attractive as a general purpose electronic structure method. 
First, the iteration of the Dyson equation produces a 
self-consistent Green function that  is independent 
of the starting point, and determines both the ground- 
and excited-state properties (quasiparticle spectra) of 
a given system on the same quantum mechanical level. 
This distinguishes sc-$GW$ from other (partially) 
self-consistent $GW$ schemes, \cite{qpscgw2004,kresse2007}
which do not lend themselves to total-energy or ionic force calculations. Moreover 
the uniqueness of the sc-$GW$
Green function facilitates an unprecedented and unbiased
assessment of the $GW$ approach, which previously was masked
by the starting-point dependence of $G_{\rm 0}W_{\rm 0}$. 
Second, self-consistency improves total and quasi-particle 
energies compared to $G_{\rm 0}W_{\rm 0}$ based on HF or DFT 
in (semi)local approximations and yields good agreement with 
high level quantum-chemical calculations and photo-emission 
data. 
Third, unlike $G_{\rm 0}W_{\rm 0}$, sc-$GW$ yields 
an associated ground-state electron density, whose 
quality is e.g. reflected in the improved description of dipole moments.
All these points 
taken together are essential for future developments in 
electronic-structure theory such as vertex functions and 
beyond $GW$ approaches.

In the $GW$ approximation the electron self-energy $\Sigma$ is defined 
in terms of the one-particle Green function $G$ and the 
screened Coulomb interaction $W$ as
\begin{equation}\label{eq:sigma}
\Sigma ({\bf r},{\bf r'},\omega) = 
i\int \frac{d\omega'}{2\pi} G({\bf r},{\bf r'},\omega+\omega')
W({\bf r},{\bf r'},\omega')e^{i\omega\eta} \quad,
\end{equation}
where $\eta$ is a positive infinitesimal, $W$  is the screened interaction
and spin-variables are omitted for simplicity.
More details about the calculation of $W$ 
are given in the supplemental material.\cite{sup}
\begin{figure}
\includegraphics[width=0.45\textwidth]{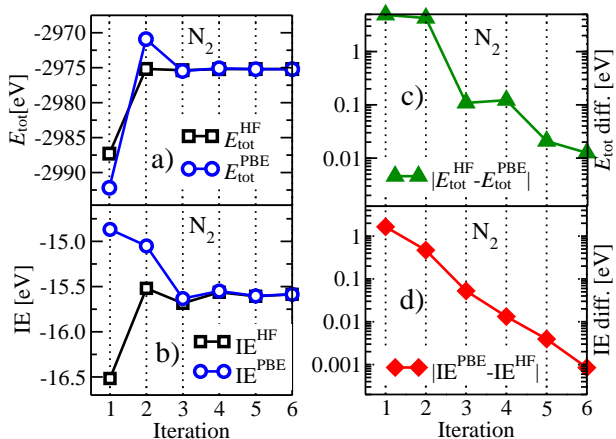}
\caption{\label{fig:start} (Color online)
Total energy (a) and ionization energy (b) of N$_2$ at each iteration of the sc-$GW$
loop for a HF and PBE input Green function, for the aug-cc-pVQZ \cite{gaussianbasis1989} basis set.
The (absolute values of the) differences arising from HF and PBE initializations vanish exponentially for both
total energy (c) and ionization energy (d).
}
\end{figure}
The Green function can in turn be expressed in terms of the self-energy 
through the Dyson equation 
\begin{equation}\label{eq:dyson}
G^{-1}=G_{\rm 0}^{-1}-[\Sigma-v_{0} + \Delta v_{\rm H}] \quad,
\end{equation}
where $G_{\rm 0}$ refers to the Green function of an
independent particle system in an effective potential $v_{\rm eff}$. $\Delta v_{\rm H}$ accounts for
changes in the Hartree potential due to density differences 
between $G_{\rm 0}$ and $G$, and $v_0$ 
is the exact-exchange operator in HF or the Kohn-Sham exchange-correlation potential in DFT.
The interdependence of Eq.~\ref{eq:sigma} and Eq.~\ref{eq:dyson}
is the very origin of the self-consistent nature of the $GW$ approach. 

It is common practice, however, to solve Eq.~\ref{eq:sigma} and Eq.~\ref{eq:dyson} 
just once, in the so-called {\it one-shot} $G_{\rm 0}W_{\rm 0}$ approximation. 
If Eq.~\ref{eq:dyson} is not solved, 
the $G_{\rm 0}W_{\rm 0}$ quasi-particle energies are given 
in first-order perturbation theory as corrections to the reference eigenvalues 
($\{\epsilon_n^{0}\}$) \cite{hybertsenlouie1986} as $\epsilon_n^{\rm QP}=\epsilon_n^{0}+
Re\left\langle \psi_n^0 \right| \Sigma(\epsilon_n^{\rm QP})- v_{0}\left|\psi_n^0 \right\rangle $.

In this work, Eqs.~\ref{eq:sigma}-\ref{eq:dyson} were solved iteratively.
Most importantly the screened interaction $W$ is also updated at each 
iteration taking into account the full 
frequency dependence of the polarizability $\chi_0$ on the imaginary axis. 
In sc-$GW$ the excitation spectrum is given by the (integrated) spectral function:
\begin{equation}\label{eq:spectrum}
A(\omega)=-1/\pi\int d{\bf r}\lim_{{\bf r'}\rightarrow{\bf r}}Im G({\bf r},{\bf r'},\omega) \quad .
\end{equation}
The ground-state density $ n({\bf r})$ also follows directly from the Green function: \cite{fetter}
\begin{equation}\label{eq:density}
    n({\bf r})=-2i G({\bf r},{\bf r},\tau=0^-) \quad.
\end{equation}
The number of particles 
can be obtained through the integration of Eq.~\ref{eq:density}. 
This permits to verify the validity of the particle number conservation law
at self-consistency (not shown for brevity), that is violated by non self-consistent 
approaches as $G_0W_0$. 

The situation is more complicated for the total energy. 
As alluded to above, for a given Green function different prescriptions exist 
to compute the associated total energy such as the Galitskii-Migdal formula,\cite{galitskii} 
the Luttinger-Ward \cite{luttingerward1960} or the Klein 
\cite{klein1961} functional. 
The latter two are variational in the sense that they are stationary at the self-consistent Green function,
and therefore might provide better total energies than the Galitskii-Migdal 
formula when evaluated with non-self-consistent Green functions. \cite{holm1999,dahlenleeuwen2005}
However, at self-consistency all three approaches are equivalent. 
Therefore, we choose the Galitskii-Migdal formula as it is 
easier to implement:
\begin{equation}
E_{\rm GM}=-{i}\int \frac{d\omega}{2\pi} Tr\left\lbrace\left[\omega+ {h}_0 \right] G(\omega)\right\rbrace + E_{\rm ion}\label{eq:migdal} \quad,
\end{equation}
where $h_0$ is the one-particle term of the many-body Hamiltonian, i.e. 
the sum of the kinetic operator and the external potential. 
Equation \ref{eq:migdal} can be rewritten using the
equation of motion for the Green function \cite{sup} as:
\begin{align}\label{eq:rewriting}
E_{\rm GM} &= -i\sum_{ij} G_{ij}(\tau=0^-)[2t_{ji} + 2v^{\rm ext}_{ji} +v^{\rm H}_{ji}+  \Sigma^{\rm x}_{ji} ]\nonumber\\
 &-i \sum_{ij} \int\frac{d\omega}{2\pi}G_{ij}(\omega)\Sigma^{\rm c}_{ji}(\omega)e^{i\omega\eta} + E_{\rm ion} \: .
\end{align}
Here $t$ denotes the kinetic-energy operator, $v^{\rm H}$ and $v^{\rm ext}$ 
the Hartree and external potential and $\Sigma^{\rm x}$ and 
$\Sigma^{\rm c}$ are the
exchange and correlation parts of the self-energy, respectively.
In Eqs.~\ref{eq:migdal} and \ref{eq:rewriting} we suppressed spin variables,
i.e. we assumed spin-degeneracy.
The trace of Eq.~\ref{eq:migdal} is expressed as sum over  
basis functions in Eq.~\ref{eq:rewriting} 
and the frequency integration is conveniently performed
along the imaginary axis. \cite{Pablo:2001}

For comparison, we also computed $G_{\rm 0}W_{\rm 0}$ total energies with different starting points. 
However, as indicated above, $G_0W_0$ total energies are {\it not} uniquely defined, because the Green 
function and the self-energy are never on the same level. If, for example, the Dyson equation is not 
solved, $G_0$ and $\Sigma_0=G_0W_0$ enter Eq.~\ref{eq:rewriting}. If the Dyson equation is solved, 
the resulting $G_1$ is still inconsistent with $\Sigma_0$.  In the following we refer to the 
combination of $G_0$ and $\Sigma_0$ in Eq.~\ref{eq:rewriting} as $G_0W_0$ total energy and denote 
the corresponding starting point with @{\it starting point}.

We have implemented sc-$GW$ in the all-electron electronic structure code 
FHI-aims.\cite{blum_etal,Xinguo/implem} Equations \ref{eq:sigma}-\ref{eq:density} and 
Eq.~\ref{eq:rewriting} are solved in a  numerical atomic orbital (NAO) basis using the 
{\it resolution of  identity} technique to treat all two-particle operators efficiently.\cite{Xinguo/implem,RI-SVS} 
All calculations are performed on the imaginary frequency axis and 
the spectral function is obtained by analytic continuation to the real frequency axis.\cite{Xinguo/implem}
The analytic continuation constitutes the only approximation of 
our implementation of the sc-$GW$ method. 
Further details of the implementation will be given elsewhere.\cite{Caruso/tocome/all}

\begin{figure}
\includegraphics[width=0.45\textwidth]{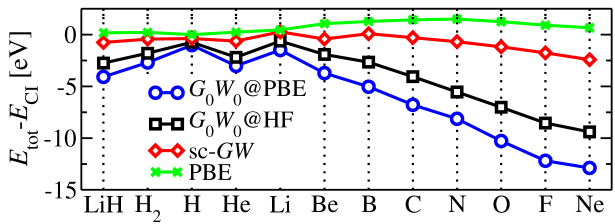}
\caption{\label{fig:etot} (Color online) Difference between Galitskii-Migdal total energies ($E_{\rm GM}$) and full 
configuration interaction values ($E_{\rm CI}$),\cite{CIatoms,LiHCI,h2ci1993}
with $E_{\rm GM}$ evaluated from sc-$GW$, 
$G_{\rm 0}W_{\rm 0}$@HF and $G_0W_0$@PBE. 
PBE total energy are included for comparison. 
The calculations were performed using the aug-cc-pV5Z basis set.\cite{gaussianbasis1989}
}
\end{figure}

In Fig.~\ref{fig:start} we demonstrate for N$_2$ that the sc-$GW$ Green function provides total 
energies (a) and vertical ionization energies (b) that are independent of the starting point.  
Figure \ref{fig:start} explicitly illustrates this point starting the self-consistency cycle 
with HF and  DFT in the Perdew, Burke, and Ernzerhof (PBE) \cite{PBE} generalized gradient approximation, 
but other initializations like the local-density approximation (LDA) or the simple Hartree approximation 
produce the same final sc-$GW$ Green function (not shown). The deviation in the Green function 
exemplified by the (absolute value of the) total energy difference (Fig.~\ref{fig:start}(c)) and 
the ionization energy difference (Fig.~\ref{fig:start}(d)) converges exponentially fast with the 
number of iterations, canceling the starting point dependence. Further tests performed on a set 
of $30$ closed shell molecules (see Fig.~\ref{fig:ip} and supplemental material) confirm this fact and demonstrate that 
sc-$GW$ provides a recipe for linking different reference systems of independent electrons 
(or non-interacting Kohn-Sham particles) to a unified interacting many-body state.

Having established the important point that the sc-$GW$ solution is 
independent of the starting point for the set of closed shell molecules studied here, 
we now turn to 
an assessment of the performance of the $GW$ approach 
for ionization potentials, electron densities and 
total energies. For elements in the first two 
rows of the periodic table (i.e. $Z=1-10$)  and
small molecules like H$_2$ and LiH accurate
reference data from configuration interaction 
(CI) calculations are available.\cite{CIatoms,h2ci1993,LiHCI} Figure \ref{fig:etot} 
reports the difference to CI values for basis set converged sc-$GW$, $G_{\rm 0}W_{\rm 0}$@PBE 
and $G_{\rm 0}W_{\rm 0}$@HF calculations. A subset of these has previously been calculated using 
sc-$GW$ \cite{stan} and our results are in excellent agreement  with the published results. 
In line with previous calculation for 
the electron gas,\cite{holm1999,kutepov,Pablo:2001} atoms and small molecules,\cite{stan}
$G_{\rm 0}W_{\rm 0}$ total energies (in various flavors) tend to be too negative.
The self-consistent 
treatment largely (but not fully) corrects this overestimation and provides total energies 
in  more satisfying agreement with full CI. The remaining overestimation provides a clear 
and unbiased quantification of the required vertex corrections in a beyond $GW$ treatment.

\begin{table}
\begin{tabular}{l   c            c  c  c}
\hline\hline
 CO & $d$ & $\nu_{\rm vib}$ & $\mu$ & $E_{\rm b}$ \\
 \hline
Exp.~\cite{expmol}  \,\,\, \,\,\,        &\,\,\, 1.128 \,\,\, &\,\,\, 2169 \,\,\, &\,\,\,  0.11 \,\,\, & 11.11  \\ 
sc-$GW$           & 1.118  & 2322  &  0.07 & 10.19 \\ 
$G_0W_0$@HF       & 1.119  & 2647  &   -   & 11.88 \\
$G_0W_0$@PBE      & 1.143  & 2322  &   -   & 12.16 \\
(EX+cRPA)@HF      & 1.116  & 2321  &   -   & 10.19 \\
(EX+cRPA)@PBE     & 1.137  & 2115  &   -   & 10.45 \\ 
PBE               & 1.135  & 2128  &  0.20 & 11.67 \\
HF                & 1.102  & 2448  & -0.13 & 7.63  \\
\hline
\hline
\end{tabular}
\caption{\label{tab:groundstate} 
Equilibrium bond length $d$, vibrational frequency 
$\nu_{\rm vib}$, dipole moment $\mu$ and binding energy $E_{\rm b}$ of the CO dimer.
Units are respectively \AA, cm$^{-1}$, Debye and eV. 
All calculation were performed with a Tier 4 basis set.}
\end{table}

For practical purposes, total energy differences are far more important than absolute total energies. 
However, for sc-$GW$ only one study has reported ground-state properties and found that sc-$GW$ 
gives lattice constants of Si and Na in good agreement with experiments.
To assess ground state properties, like the equilibrium atomic structure, would in principle require atomic 
forces (i.e. derivatives of the total energy with respect to atomic coordinates), which are presently 
not available for sc-$GW$. For diatomic molecules, however, structural properties such as vibrational 
frequencies, bond lengths and binding energies can be determined directly from the potential 
energy curve. Other ground state properties, e.g. dipole moments, can be inferred directly  from the 
electron density. For brevity, we only present the case of CO here and 
refer to a future publication for a more detailed 
discussion of ground state properties in sc-$GW$. \cite{Caruso/tocome/all}

\begin{figure*}
\includegraphics[width=0.77\textwidth]{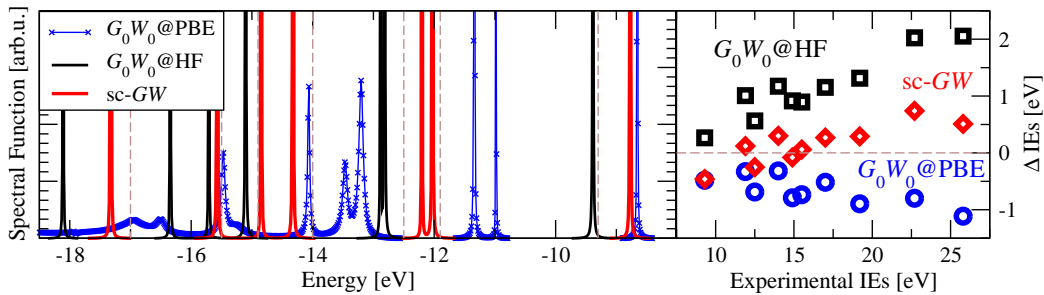}
\caption{\label{fig:spectrum} (Color online)
Left panel: The spectral function of benzene calculated with 
a Tier 2 basis set. Vertical dashed lines are located at 
experimental vertical ionization energies from Ref.~\onlinecite{expbenzene}. 
Right panel: comparison of experimental \cite{expbenzene}
and theoretical vertical ionization energies (VIEs) extracted from 
the spectral function of benzene for sc-$GW$, 
$G_{\rm 0}W_{\rm 0}$@HF and $G_0W_0$@PBE.}
\end{figure*}

In Table \ref{tab:groundstate} we report the experimental  values for the bond length $d$, vibrational frequency 
$\nu_{\rm vib}$, dipole moment $\mu$ and binding energy $E_{\rm b}$ of CO  \cite{expmol} together with the 
theoretical values obtained from several perturbative and non-perturbative approaches. DFT in the 
exact-exchange plus correlation in the random-phase approximation (EX+cRPA) 
based on PBE is remarkably accurate for the bond 
length and vibrational frequency of CO.\cite{Ren/Rinke/Scheffler:2009} However, like $G_0W_0$, 
EX+cRPA exhibits a considerable starting point dependence  and gives no direct access to dipole moments. 
In sc-$GW$, the quality of the new density, obtained through Eq.~\ref{eq:density}, is manifested in the 
improved dipole moment, which is in much better agreement with the experimental value than in PBE and HF. 
Additional information on the quality of the sc-$GW$ electron density is reported in the supplemental 
material.\cite{sup}
The vibrational frequency, on the other hand, is overestimated and not substantially different from the 
perturbative $G_0W_0$ values. Self-consistency over-corrects the overestimation of the $G_0W_0$ binding 
energy, resulting in an underestimation of about 1 eV for $E_{\rm b}$ compared to experiment.  
Similarly, the sc-$GW$ bond length is slightly too small and is close to $G_0W_0$@HF. This assessment 
of the $GW$ approach for ground state properties, facilitated by sc-$GW$, clearly indicates where 
future challenges in going beyond $GW$ lie. 

Finally, we turn to the description of spectral properties.
For the homogeneous electron gas (HEG) Holm 
and von Barth first reported a deterioration of the spectral properties 
\cite{holmvonbarth1998} in sc-$GW$ compared to $G_{\rm 0}W_{\rm 0}$@LDA. 
For the spectra of simple solids like silicon and sodium, controversy then arose with 
some authors advocating self-consistency \cite{eguiluz2002} and others dismissing it.\cite{eguiluz1998,kutepov} Part of this controversy can be traced back to convergence 
difficulties in the early all-electron calculations,\cite{Friedrich/etal:2006/all} while 
the influence of the pseudopotential approximation in $GW$ turned out to be larger than 
initially anticipated.\cite{GomezAbal08/all}

To test the quality of the sc-$GW$ spectra we 
chose the benzene molecule as a benchmark, 
for which the sc-$GW$ spectral
function in Fig.~\ref{fig:spectrum} is compared  to the
$G_{\rm 0}W_{\rm 0}$@HF and $G_0W_0$@PBE ones calculated using Eq.~\ref{eq:spectrum}.
The vertical ionization energies (VIEs) shown in the right panel of
Fig.~\ref{fig:spectrum} correspond to the  peak positions in the
spectral function.
All the peaks reported in the left panel of Fig. 3 
correspond to occupied quasi-particle states and the 
associated energy can be directly related to ionization 
energies as measured in photoemission spectroscopy.
The $G_{\rm 0}W_{\rm 0}$ quasi-particle energies -- reported in the right panel of Fig.~\ref{fig:spectrum} -- 
depend strongly on the starting point: 
HF-(PBE-)based $G_{\rm 0}W_{\rm 0}$  has a tendency 
to overestimate (underestimate) VIEs.
The deviation between $G_{\rm 0}W_{\rm 0}$@HF and $G_0W_0$@PBE is $\simeq 0.5 $ eV for 
the first ionization energy and can be as large as $\simeq 3 $ eV for lower 
lying quasi-particle states. 
Furthermore, due to overscreening $G_{\rm 0}W_{\rm 0}$@PBE yields a large broadening 
(i.e. short lifetimes) for quasi-particle peaks below $-12$ eV. 
We emphasize that those peaks are not  
plasmon satellites, but quasi-particle states with a short lifetime. 
The short lifetime arises from the small HOMO-LUMO gap in PBE that 
allows quasi-particle states to decay
through the creation of electron-hole pairs \cite{marinirubio2004}.
At self-consistency, the quasi-particle energies 
are uniquely defined, the systematic (over)underestimation of $G_{\rm 0}W_{\rm 0}$ 
calculations is considerably reduced and the resulting
quasi-particle energies are in better agreement with photoemission data.\cite{expbenzene}

We further assessed the quality of sc-$GW$ quasi-particle energies for the set of 30 molecules 
calculated by Rostgaard {\it et al.}.\cite{thygesen} For brevity our results are summarized in 
Fig.~\ref{fig:ip} and we refer to the supplemental material for the actual numerical values.\cite{sup}
In Ref.~\onlinecite{thygesen} sc-$GW$ was based on the frozen-core approximation, 
whereas in our work core electrons were also treated fully self-consistently.
Core-valence coupling is therefore included in our implementation 
and is likely responsible for the deviation of 0.1-0.5 eV in the first VIEs between our
and Rostgaard {\it et al.'s} sc-$GW$ calculations.
As for benzene, $G_{\rm 0}W_{\rm 0}$@HF tends to overestimate VIEs, while $G_0W_0$@PBE underestimates. 
sc-$GW$ also slightly underestimates the VIEs, but gives an average deviation of only  $2\%$ compared to 
$6\%$ in $G_0W_0$@PBE and $4\%$ in $G_0W_0$@HF. PBE and HF present two extreme starting points. 
In PBE the gap between the highest occupied molecular orbital (HOMO) and the lowest unoccupied 
molecular orbital (LUMO) is severely underestimated, while in HF it is considerably overestimated. 
This in part explains the behavior of  $G_0W_0$@PBE and $G_0W_0$@HF. Since the screening strength 
is inversely proportional to the HOMO-LUMO gap, $G_0W_0$@PBE overscreens and $G_0W_0$@HF underscreens. 
By tuning the fraction of exact exchange in the ground states, as e.g. in hybrid functionals, 
the deviation between $G_0W_0$ and experiment could be further reduced for this data set.  
However, this procedure is neither predictive, nor universal or transferable, because different 
systems will require a different amount of exact exchange. To really assess the quality of the 
$GW$ approximation, self-consistency is therefore indispensable. 
From this we conclude that sc-$GW$ systematically improves the spectral properties of 
the systems considered here as compared to perturbative $GW$. 
More work is needed to investigate the quality of sc-$GW$ for 
a wider range of systems and materials, including 
transition metals or rare earth elements were 
correlations are stronger.

\begin{figure}
\includegraphics[width=80mm]{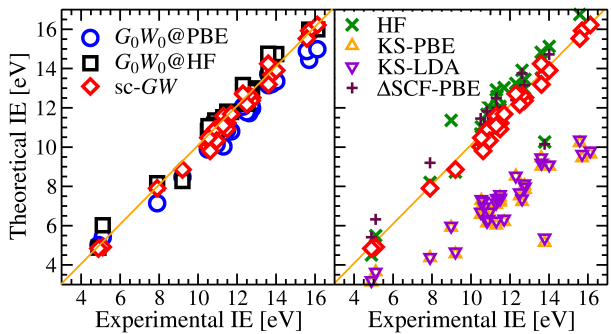}
\caption{\label{fig:ip} (Color online) 
First vertical ionization energy (VIE) for $30$ closed-shell 
molecules composed of 2 to 8 atoms. 
Experimental values are taken from Ref.~\onlinecite{expmol}.
Results from PBE total energy differences ($\Delta$ SCF-PBE) are included for comparison.
}
\end{figure}

In summary,  we have demonstrated that sc-$GW$ is 
independent of the starting point for closed shell 
molecules. Self-consistency improves the total 
energy and the spectral properties of the test 
sets compared to $G_0W_0$ based on HF or PBE, whereas structural properties worsen compared to EX+cRPA.
Moreover, the sc-$GW$ electron densities 
improve the description of the dipole moment of CO. The sc-$GW$ approach therefore provides a unified theory 
for the  electronic ground- and excited-state properties 
of  many-body systems. Most importantly, sc-$GW$  gives 
unambiguous reference data that is essential for 
developing vertex corrections for (bio)molecules, 
nanostructures and extended systems, in particular 
for the challenging class of ``strongly correlated'' materials.

AR acknowledges financial support from MEC (FIS2011-65702-C02-01),  Grupos Consolidados UPV/EHU del Gobierno Vasco (IT-319-07), and the European Research Council (ERC-2010-AdG -No. 267374).


%

\end{document}